\documentclass[12pt,epsfig]{article}
\usepackage[xdvi]{graphicx}

\def\fund{  \> {\vcenter  {\vbox
               {\hrule height.6pt
                \hbox {\vrule width.6pt  height5pt
                      \kern5pt
                      \vrule width.6pt  height5pt}
                \hrule height.6pt}
                         }
               }
            \>\>  }

\def\sym{  \>\underbrace{ {\vcenter  {\vbox
              {\hrule height.6pt
               \hbox {\vrule width.6pt  height5pt
                      \kern5pt
                      \vrule width.6pt  height5pt
                      \kern5pt
                      \vrule width.6pt  height5pt
                      \kern5pt
                      \vrule width.6pt  height5pt
                      \kern5pt
                      \vrule width.6pt  height5pt}
               \hrule height.6pt}
                         }
               }}_{r}
            \>\>  }

\newcommand{\beq}{\begin{equation}}
\newcommand{\eeq}{\end{equation}}
\newcommand{\bea}{\begin{eqnarray}}
\newcommand{\eea}{\end{eqnarray}}

\def\e2sig{e^{-2r\sigma}}

\begin{document}

\newcommand{\mpl}{M_{\mathrm{Pl}}}
\setlength{\baselineskip}{18pt}

\begin{titlepage}

\vspace*{1.0cm}
\begin{center}
{\Large\bf Calculable Violation of Gauge-Yukawa Universality \\ 
\vspace*{5mm} 
and Top Quark Mass in the Gauge-Higgs Unification} 
\end{center}
\vspace{25mm}

\begin{center}
{\large
C. S. Lim
and Nobuhito Maru$^*$}
\end{center}
\vspace{1cm}
\centerline{{\it Department of Physics, Kobe University,
Kobe 657-8501, Japan}} 
\vspace*{2mm}
\centerline{{\it $^*$Department of Physics, Chuo University,
Tokyo 112-8551, Japan}}
%
%
\vspace{2cm}
\centerline{\large\bf Abstract}
\vspace{0.5cm}
We find that the one-loop correction to 
the ratio of Yukawa coupling and gauge coupling in the gauge-Higgs unification, 
``gauge-Yukawa universality violation", is finite and calculable in any space-time dimension. 
Applying this result to the ratio of top quark and W-boson masses, 
we show that the order one correction required to generate a viable top quark mass 
is indeed possible if the fermion embedding top quark belongs to the large dimensional representation 
of the gauge group and a vacuum expectation value of Higgs scalar field is very small 
comparing to the compactification scale.

\end{titlepage}

Gauge-Higgs unification \cite{Manton, Fairlie, Hosotani} is one of the attractive scenarios 
solving the hierarchy problem without invoking supersymmetry. 
In this scenario, 
Higgs doublet in the Standard Model (SM) is identified with 
the extra spatial components of the higher dimensional gauge fields. 
One of the remarkable features is that 
quantum corrections to Higgs mass is insensitive 
to the cutoff scale of the theory and calculable 
regardless of the nonrenormalizability of higher dimensional gauge theory. 
The reason is that the Higgs mass term as a local operator is forbidden 
by the higher dimensional gauge invariance. 
Then, the finite mass term is generated radiatively and 
expressed by the Wilson line phase as a non-local operator. 
This fact has opened up a new avenue to the solution of the hierarchy problem \cite{HIL}. 
Since then, much attention has been paid to the gauge-Higgs unification and 
many interesting works have been done from various points of view 
\cite{KLY}-\cite{LHC}. 

The finiteness of Higgs mass has been studied and verified in various models 
and in various types of compactification at one-loop level\footnote{For the case of gravity-gauge-Higgs unification, 
see \cite{HLM}} \cite{ABQ}-\cite{LMH} and even at two loop level \cite{2-loop}. 
It is natural to ask whether any other finite physical observables exist in the gauge-Higgs unification. 
The naive guess is that such observables are in the gauge-Higgs sector of the theory if they ever exist. 
The present authors studied the structure of divergences for S and T parameters 
in the gauge-Higgs unification since such parameters are described 
by higher dimensional gauge invariant operators with respect to gauge and Higgs fields, 
and are expected to be finite by virtue of the higher dimensional gauge symmetry. 
The result is that both parameters are divergent (convergent) more than (in) five dimensions 
as expected from the power counting argument.  
However, a nontrivial prediction  we have found, specific to the gauge-Higgs unification, is 
that some linear combination of S and T parameters is finite even in six dimensions \cite{LM}. 
One of the authors (N.M.) also has shown that the gluon fusion amplitude and 
the amplitude of two photon decay of Higgs boson, which are very important processes at LHC, 
are finite in any space-time dimension in the gauge-Higgs unification \cite{NM}. 
This is a new and only known calculable physical observable 
other than the Higgs mass.\footnote{In a toy model of QED compactified on a circle, 
the anomalous magnetic moment was shown to be finite in arbitrary space-time dimensions \cite{ALM}. 
Recently, the cancellation mechanism of the ultraviolet (UV) divergence in the magnetic moment 
was further clarified in a realistic $SU(3)$ model on an orbifold $S^1/Z_2$ \cite{ALM2}.}

Although the gauge-Higgs unification is very predictive 
in the gauge-Higgs sector of the standard model 
as mentioned above, 
the matter sector is too restrictive to generate a desirable flavor structure 
since Yukawa coupling is given by the gauge coupling, to start with. 
This immediately leads to the fact that fermion masses become W-boson mass 
and Yukawa hierarchy cannot be explained, 
unless some suitable mechanism is adopted. 
Obtaining light fermion masses is easily realized by introducing the $Z_2$-odd bulk mass 
because the zero mode wave functions for fermions with different chiralities 
are localized at different fixed points along the extra dimension, 
which naturally yields a small Yukawa coupling 
due to the small overlap integral of the zero mode wave functions.  
However, getting the top quark mass is nontrivial task since we need an enhancement factor of roughly 2, 
as $m_t \simeq 2 m_W$. 
In flat space gauge-Higgs unification model, 
this enhancement factor can be obtained from the group theoretical factor of 
large dimensional representation which a bulk fermion embedding top quark belongs to \cite{CCP}. 
In warped space case, it is known that the enhancement factor comes 
from the product of curvature scale and compactification radius \cite{HM}.

In this paper, we propose an alternative mechanism to generate a viable top quark mass 
by taking into account one-loop corrections to Yukawa coupling 
in the flat space gauge-Higgs unification. 
Naively thinking, this seems to be clearly impossible because the loop corrections are always suppressed. 
However, this is not necessarily the case in the case of gauge-Higgs unification. 
As will be shown later, the one-loop correction effects have additional factor of 
Dynkin index for the representation which the matter fermion belongs to
and a logarhythmic factor of Higgs vacuum expectation value (VEV) other than the one-loop factor. 
If we consider the fermion belonging to large dimensional representation, 
we can have a large Dynkin index. 
We further note that the logarhythmic factor of Higgs VEV is likely to be large 
since the Higgs VEV should be tiny compared to the compactification scale, 
typically around ${\cal O}(10^{-2})$, 
to realize the correct pattern of electroweak symmetry breaking 
and obtain a Higgs mass satisfying the experimental data. 
Combining these effects, we can expect that the one-loop correction to Yukawa coupling becomes ${\cal O}(1)$. 

Quantum correction to Yukawa coupling is generally a cutoff scale dependent, 
especially in the nonrenormalizable higher dimensional theories, 
and is independent of the quantum correction to the gauge coupling. 
Thus, the quantum correction seems to have no definite prediction. 
In the gauge-Higgs unification, however, Yukawa and gauge coupling are identical, to start with, 
being described by the same covariant derivative, {\em i.e.} ``gauge-Yukawa universality" holds. 
Hence, even if the universality is violated at the quantum level, the violation should be finite and calculable. 
It is interesting to note that a similar situation happens to MSSM, 
where Higgs self-coupling is provided by the gauge interaction, D-term, 
and the deviation of the Higgs mass from the gauge boson masses is finite, even at the quantum level \cite{Higgsmassbound}. 
According to this line of argument, 
we calculate here one-loop corrections to the ratio of Yukawa coupling and the gauge coupling, 
which will be shown to be independent of the cutoff scale of the theory, namely calculable and finite. 
Since Yukawa coupling is provided by a part of the gauge coupling $g \bar{\psi} A_M \Gamma^M \psi$, 
the renormalized Yukawa coupling is obtained 
by taking into account the wave function renormalization factors 
of extra component of the gauge field $A_y = \sqrt{Z_y} A_y^{{\rm bare}}$, 
a fermion $\psi = \sqrt{Z_{\psi}} \psi^{{\rm bare}}$, and the vertex correction $Z_{A_y \psi \psi}$. 
\bea
Y^{{\rm ren}} = \frac{Z_{A_y \psi \psi}}{Z_\psi \sqrt{Z_y}} Y = \frac{1}{\sqrt{Z_y}} Y
\label{top}
\eea
where $Y$ and $Y^{{\rm ren}}$ are the bare and renormalized Yukawa couplings 
and we made use of Ward identity $Z_{A_y \psi \psi}=Z_{\psi}$ to arrive at the final expression. 
On the other hand, it is well known that 
the renormalized gauge coupling $g^{{\rm ren}}$ is calculated from the vacuum polarization of gauge field, 
\bea
g^{{\rm ren}} = \frac{1}{\sqrt{Z_\mu}} g 
\label{Wmass}
\eea
where $g$ is the bare gauge coupling and 
$Z_\mu$ denotes the wave function renormalization factor for the gauge field $A_\mu$, 
namely $A_\mu = \sqrt{Z_\mu} A_\mu^{{\rm bare}}$. 
Taking the ratio of (\ref{top}) and (\ref{Wmass}) 
using the gauge-Yukawa universality $Y=g$, 
we find 
\bea
\frac{Y^{{\rm ren}}}{g^{{\rm ren}}} = \sqrt{\frac{Z_\mu}{Z_y}}. 
\label{ratio}
\eea
\begin{figure}[h] 
\begin{center}
  \includegraphics[width=8cm]{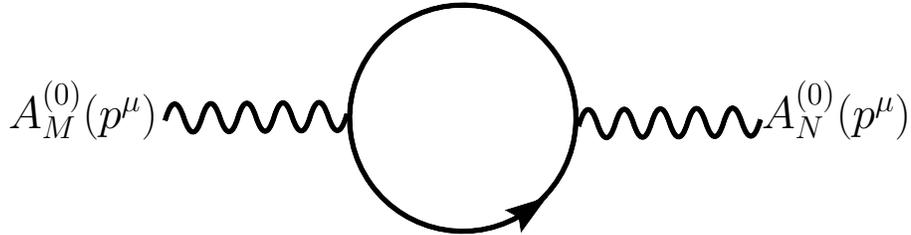}
  \put(-285,45){{\Large $A_M^{(0)}(p^\mu)$}}
  \put(0,45){{\Large $A_N^{(0)}(p^\mu)$}}
 \end{center}
 \vspace*{-0.5cm}
\caption{
Vacuum polarization diagram by fermion loop for the zero mode. 
$p^\mu$ denotes $D$-dimensional external momentum. 
} 
\label{vp}
\end{figure}
Thus, we have only to calculate the vacuum polarization diagram, 
defined as $\Pi_{MN}(p^2)$, shown in Fig. \ref{vp}. 
To be precise, we are interested in the difference between $\Pi_{\mu\nu}$ and $\Pi_{yy}$. 
Note that the nonzero KK external momenta are set to be zero 
since we are only interested in wave function renormalization factors of the zero modes. 
Each wave function renormalization factors in (\ref{ratio}) are divergent, 
but the ratio is expected to be finite from the argument above and 
also simply from the higher dimensional Lorentz invariance.\footnote{Similar ratio appeared 
in the calculation of Higgs mass at two-loop level \cite{2-loop}.} 
In fact, the wave function renormalization factor $Z$ for the local operator $ZF_{MN}F^{MN}$ is universal. 
Thus, we can make UV-insensitive prediction 
for the one-loop correction to top quark mass by making use of this ratio. 

In this paper, we take a $(D+1)$-dimensional $SU(3)$ gauge theory 
with a triplet fermion compactified on $S^1/Z_2$. 
As will be seen later, the triplet fermion contains a doublet top quark $t_L$ but not a singlet one $t_R$ and 
a fermion belonging to large dimensional representation accommodating $t_R$ 
is necessary to obtain a viable top quark mass. 
The fermion of large dimensional representation also helps to get a realistic top quark mass. 
The contribution by such a fermion in the large dimensional representation to top quark mass 
can be reduced to the contribution of a triplet fermion multiplied by an additional group factor, 
namely Dynkin index. Therefore, the calculation throughout this paper is carried out by using the triplet fermion.  
The $SU(3)$ symmetry is broken to $SU(2) \times U(1)$ 
by the orbifolding $S^1/Z_2$ and adopting a non-trivial $Z_{2}$ parity assignment 
for the members of an irreducible representation of $SU(3)$, as stated below. 
The remaining gauge symmetry $SU(2) \times U(1)$ is supposed to be broken 
by the VEV of the zero-mode of $A_{y}$, 
the extra space component of the gauge field behaving as the Higgs doublet, 
through the Hosotani-mechanism \cite{Hosotani}, 
though we do not address the question how the VEV is obtained 
by minimizing the loop-induced effective potential for $A_y$ \cite{KLY}.      

The lagrangian is simply given by 
\bea
{\cal L} = -\frac{1}{2} \mbox{Tr}  (F_{MN}F^{MN}) 
+ i\bar{\Psi}D\!\!\!\!/ \Psi
\label{lagrangian}
\eea
where $\Gamma^M=(\gamma^\mu, i \gamma^y)$, 
\bea
F_{MN} &=& \partial_M A_N - \partial_N A_M -i g_{D+1} [A_M, A_N]~(M,N = 0,1,2,3,\cdots, D), \\
D\!\!\!\!/ &=& \Gamma^M (\partial_M -ig_{D+1} A_M) \ \ 
\left(A_{M} = A_{M}^{a} \frac{\lambda^{a}}{2} \ 
( \lambda^{a}: \mbox{Gell-Mann matrices}) \right),  \\
\Psi &=& (\psi_1, \psi_2, \psi_3)^T.
\eea
The periodic boundary conditions are imposed along $S^1$ for all fields and 
the non-trivial $Z_2$ parities are assigned for each field  as follows, 
\bea 
\label{z2parity} 
A_\mu = 
\left(
\begin{array}{ccc}
(+,+) & (+,+) & (-,-) \\
(+,+) & (+,+) & (-,-) \\
(-,-) & (-,-) & (+,+) 
\end{array}
\right), \ \ 
A_y = 
\left(
\begin{array}{ccc}
(-,-) & (-,-) & (+,+) \\
(-,-) & (-,-) & (+,+) \\
(+,+) &(+,+) & (-,-)
\end{array}
\right), 
\eea
\bea 
\label{fermizero} 
\Psi = 
\left(
\begin{array}{cc}
\psi_{1L}(+,+) + \psi_{1R}(-, -) \\
\psi_{2L}(+,+) + \psi_{2R}(-, -) \\
\psi_{3L}(-,-) + \psi_{3R}(+, +) \\
\end{array}
\right),
\eea
where $(+,+)$ means that $Z_2$ parities are even at the fixed points $y=0$ 
and $y = \pi R$, for instance. $y$ is the compactified space coordinate and 
$R$ is the compactification radius. 
$\psi_{1L} \equiv \frac{1}{2}(1-\gamma^y)\psi_1$, etc. 
A remarkable feature of this manipulation of ``orbifolding" is that 
in the gauge-Higgs sector, 
exactly what we need for the formation of the standard model is obtained at low energies; 
one can see that $SU(3)$ is broken to $SU(2)_{L} \times U(1)_{Y}$ and the Higgs doublet 
$\phi = (\phi^{+}, \phi_{0})^{t}$ emerges. 
Namely the zero mode of the gauge-Higgs sector takes the form, 
\bea
A^{(0)}_{\mu} = \frac{1}{2}
\left(
\begin{array}{ccc}
\frac{2}{\sqrt{3}}\gamma_\mu & \sqrt{2} W^{+}_{\mu} & 0 \\
\sqrt{2} W^{-}_{\mu} & - \frac{1}{\sqrt{3}}(\gamma_\mu + \sqrt{3}Z_\mu) & 0 \\
0& 0 & -\frac{1}{\sqrt{3}}(\gamma_\mu - \sqrt{3}Z_\mu)   
\end{array}
\right) , \ \ 
A_y^{(0)} = \frac{1}{\sqrt{2}}
\left(
\begin{array}{ccc}
0 & 0 & \phi^{+} \\
0 & 0 & \phi^{0} \\
\phi^{-} & \phi^{0\ast} & 0 
\end{array}
\right),   
\eea
with $W_\mu^{3}, \ W_{\mu}^{\pm}$, $B_\mu$ being the $SU(2)_{L}, U(1)_{Y}$ gauge fields, 
respectively, while in the zero-mode of the triplet fermion $t_{R}$ is lacking, 
\bea
\Psi^{(0)} = 
\left(
\begin{array}{cc}
t_{L} \\
b_{L} \\
b_{R} \\
\end{array}
\right). 
\eea
The VEV to break $SU(2)_{L} \times U(1)_{Y}$ is written as 
\beq 
\langle A_{y} \rangle = 
\frac{v}{2} \ \lambda_{6} \ \ (\langle \phi^{0} \rangle = \frac{v}{\sqrt{2}}). 
\eeq  

Depending on these boundary conditions, 
KK mode expansions for the gauge fields and fermions are carried out as follows. 
\bea
A_{\mu,y}^{(+,+)}(x,y) &=& \frac{1}{\sqrt{2 \pi R}} 
\left[
A_{\mu,y}^{(0)}(x) + \sqrt{2} \sum_{n=1}^\infty A_{\mu,y}^{(n)}(x) 
\cos \left( \frac{ny}{R} \right)
\right], \\
A_{\mu,y}^{(-,-)}(x,y) &=& \frac{1}{\sqrt{\pi R}} 
\sum_{n=1}^\infty A_{\mu,y}^{(n)}(x) 
\sin \left( \frac{ny}{R} \right), \\
\psi_{1L, 2L, 3R}^{(+,+)}(x,y) &=& \frac{1}{\sqrt{2 \pi R}} 
\left[
\psi_{1L, 2L, 3R}^{(0)}(x) 
+ \sqrt{2} \sum_{n=1}^\infty \psi_{1L,2L,3R}^{(n)}(x) \cos \left( \frac{ny}{R} \right)
\right], \\
\psi_{3L,1R,2R}^{(-,-)}(x,y) &=& \ \frac{i}{\sqrt{\pi R}} 
\sum_{n=1}^\infty \psi_{3L,1R,2R}^{(n)}(x) \sin \left( \frac{ny}{R} \right). 
\label{KK}
\eea 
For the calculation of one-loop corrections due to the Yukawa and gauge couplings of fermions,     
only the term containing fermions, 
${\cal L}_{{\rm fermion}}  = i\bar{\Psi}D\!\!\!\!/ \Psi$, 
in the lagrangian (\ref{lagrangian}) is enough to consider. 
Substituting the above KK expansions for the fermion and the zero-modes for the gauge-Higgs bosons in the term 
and integrating over the extra space coordinate $y$,  
we obtain a 4D effective Lagrangian:  
\bea
&&{\cal L}_{{\rm fermion}}^{(4D)}  
= \sum_{n=1}^{\infty} \left\{  (\bar{\psi}_1^{(n)}, 
\bar{\tilde{\psi}}_2^{(n)}, \bar{\tilde{\psi}}_3^{(n)}) 
\right. \nonumber \\ 
&& \left. \times \left(
\begin{array}{ccc}
i \gamma^{\mu} \partial_{\mu} - m_{n} & 0 & 0 \\
0 & i \gamma^{\mu} \partial_{\mu} -(m_{n} + m ) & 0 \\
0& 0 &i \gamma^{\mu} \partial_{\mu} -(m_{n} - m)  
\end{array}
\right)
\left(
\begin{array}{c} 
\psi_1^{(n)} \\
\tilde{\psi}_2^{(n)} \\
\tilde{\psi}_3^{(n)}
\end{array}
\right) \right. \nonumber \\ 
&& \left. 
+\frac{g_D}{2} (\bar{\psi}_1^{(n)}, \bar{\tilde{\psi}}_2^{(n)}, 
\bar{\tilde{\psi}}_3^{(n)})  
\left(
\begin{array}{ccc}
\frac{2}{\sqrt{3}} \gamma_\mu & W^{+}_{\mu} & -W^{+}_{\mu} \\
W^{-}_{\mu} & - \frac{1}{\sqrt{3}} \gamma_\mu & Z_{\mu}  \\
-W^{-}_{\mu} & Z_{\mu} & - \frac{1}{\sqrt{3}} \gamma_\mu 
\end{array}
\right) 
\gamma^{\mu} 
\left(
\begin{array}{c}
\psi_1^{(n)} \\
\tilde{\psi}_2^{(n)} \\
\tilde{\psi}_3^{(n)}
\end{array}
\right) \right\} \nonumber  \\ 
&& \left. 
+\frac{g_D}{2} (\bar{\psi}_1^{(n)}, \bar{\tilde{\psi}}_2^{(n)}, 
\bar{\tilde{\psi}}_3^{(n)})  
\left(
\begin{array}{ccc}
0 & \phi^{+} & \phi^{+} \\
\phi^{-} & 
0 & - i \phi^0  \\
\phi^{-} & i \phi^0 & 0
\end{array}
\right) 
\left(
\begin{array}{c}
\psi_1^{(n)} \\
\tilde{\psi}_2^{(n)} \\
\tilde{\psi}_3^{(n)}
\end{array}
\right) \right\} \nonumber  \\
&&  + i \bar{t}_{L} \gamma^{\mu} \partial_{\mu} t_{L} 
+  \bar{b} (i \gamma^{\mu} \partial_{\mu} - m) b 
+ \frac{\sqrt{3}g_D}{6} (\bar{t}\gamma_{\mu} L t 
+ \bar{b}\gamma_{\mu} L b -2\bar{b}\gamma_{\mu} R b) B^{\mu}
\nonumber \\ 
&& +\frac{g_D}{\sqrt{2}} (\bar{t}\gamma_{\mu} L b  W^{+\mu} 
+ \bar{b}\gamma_{\mu} L t W^{-\mu}) 
+ \frac{g_D}{2} (\bar{t}\gamma_{\mu} L t 
- \bar{b}\gamma_{\mu} L b) W_{3}^{\mu} 
\label{4Deff}
\eea
where $m_{n} = \frac{n}{R}$. 
$g_D = \frac{g_{D+1}}{\sqrt{2\pi R}}$ is the $D$-dimensional gauge coupling 
and $m = \frac{g_D v}{2} (= m_{W})$ is the bottom quark mass $m_{b}$.\footnote{Top Yukawa coupling is not generated 
in the case of triplet fermion. We will later consider the fermion 
in the large dimensional representation inducing top Yukawa coupling, such as ${\bf 15}$.} 
In deriving the 4D effective Lagrangian (\ref{4Deff}), 
a chiral rotation 
\beq  
\psi_{1,2,3} \ \to \ e^{-i\frac{\pi}{4}\gamma^y} \psi_{1,2,3}  
\eeq
has been made in order to get rid of $i \gamma^{y}$.  
We easily see that the non-zero KK modes in the mass eigenstates 
$\tilde{\psi}_{2}^{(n)}, \ \tilde{\psi}_{3}^{(n)}$ after the electroweak symmetry breaking are obtained as,   
\bea 
\pmatrix{ 
\psi_{1}^{(n)} \cr 
\tilde{\psi}_{2}^{(n)} \cr 
\tilde{\psi}_{3}^{(n)} \cr 
} 
= {\cal O} 
\pmatrix{ 
\psi_{1}^{(n)} \cr 
\psi_{2}^{(n)} \cr 
\psi_{3}^{(n)} \cr 
}, \ \ \  
{\cal O} =\frac{1}{\sqrt{2}}
\left(
\begin{array}{ccc}
\sqrt{2} & 0 & 0 \\
0 & 1 & 1 \\
0 & -1 & 1 
\end{array}
\right). 
\eea
The relevant Feynman rules for our calculation can be readily 
read off from this lagrangian.

First, we compute $\mu\nu$ components ({\em i.e.} $D$-dimensional components) 
of the vacuum polarization tensor $\Pi_{MN}(p^2)$ 
where a zero mode and nonzero KK modes of triple fermions are running in the loop. 
As an example, we consider the polarization tensor $\Pi_{\mu\nu}$ of photon $\gamma_\mu$. 
The result is 
\bea
\Pi_{\mu\nu}(p^2) 
= 
\frac{2^{[D/2]}g_D^2}{2(4\pi)^{D/2}} 
(p_\mu p_\nu - p^2 g_{\mu\nu}) 
\int_0^\infty dt t^{1-D/2} 
\sum_{n=-\infty}^\infty R \sqrt{\frac{\pi}{t} }e^{-\frac{(\pi Rn)^2}{t} - 2\pi i na}   
\label{mnvp}
\eea
where a dimensionless constant $a$ is defined as $a \equiv m_W R$. 
In the above calculation, Poisson resummation formulae are applied. 
\bea
\sum_{n=-\infty}^\infty e^{-(\frac{n+a}{R})^2 t} &=& \sum_{m=-\infty}^\infty R \sqrt{\frac{\pi}{t}} e^{-(\pi Rm)^2/t -2\pi i ma}, 
\label{Poisson1}\\
\sum_{n=-\infty}^\infty \left(\frac{n+a}{R} \right)^2e^{-(\frac{n+a}{R})^2 t} 
&=& \sum_{m=-\infty}^\infty R \left(\frac{1}{2} \sqrt{\frac{\pi}{t^3}} - \sqrt{\frac{\pi}{t^5}} (\pi Rm)^2\right) 
e^{-(\pi Rm)^2/t -2\pi i ma}.  
\label{Poisson2}
\eea
Note that only the relevant terms of order ${\cal O}(p^2)$ for the wave function renormalization factor are extracted in (\ref{mnvp}). 
The divergence appears only in the zero winding mode ($n=0$ mode after Poisson resummation), 
we thus obtain the divergent and finite part of the wave function renormalization factor as 
\bea
\Pi_{\mu\nu}^{{\rm div}}(p^2) &=& \frac{2^{[D/2]}g_D^2}{2(4\pi)^{D/2}} R \sqrt{\pi} (p_\mu p_\nu - p^2 g_{\mu\nu}) 
\int_0^\infty dt t^{(1-D)/2} 
\label{mndiv}\\
\Pi_{\mu\nu}^{{\rm finite}}(p^2) &=& 
\frac{2^{[D/2]}g_D^2}{2(4\pi)^{D/2}} (p_\mu p_\nu - p^2 g_{\mu\nu}) \int_0^\infty dt t^{1-D/2} 
\sum_{n=1}^\infty R \sqrt{\frac{\pi}{t}} e^{-\frac{(\pi Rn)^2}{t}} 2 \cos(2\pi na) 
\nonumber \\
&=& \frac{2^{[D/2]}g_D^2}{2(4\pi)^{D/2}} 2 R \sqrt{\pi} \Gamma \left(\frac{D-3}{2} \right) 
(p_\mu p_\nu - p^2 g_{\mu\nu})
\sum_{n=1}^\infty \frac{\cos(2\pi na)}{(\pi Rn)^{D-3}}.  
\label{mnfinite}
\eea
To see the violation of gauge-Yukawa universality, 
we next calculate the $yy$ component of the vacuum polarization tensor. 
As a matter of fact, 
the $A_y$ partner of the photon, say $\gamma_y$ does not have a zero mode. 
We, however, expect that at least the UV-divergence due to the quantum correction to 
an SU(3) invariant local operator is common, irrespectively of the choice of the gauge generator. 
So, $\gamma_y$ is expected to mimic, say $\phi^0$. 
The result reads as
\bea
\Pi_{yy}(p^2) 
&=& 
\frac{2^{[D/2]}g_D^2}{2(4\pi)^{D/2}} p^2 
\sum_{n=-\infty}^\infty \int_0^\infty dt t^{1-D/2} R\sqrt{\frac{\pi}{t}} 
\left(1+ \frac{(\pi Rn)^2}{t} \right) 
e^{-\frac{(\pi Rn)^2}{t} - 2\pi i na}. 
\label{yyvp}
\eea
The divergent and finite part are found, 
\bea
\Pi_{yy}^{{\rm div}}(p^2) &=& \frac{2^{[D/2]}g_D^2}{2(4\pi)^{D/2}} p^2 
\int_0^\infty dt t^{(1-D)/2} R\sqrt{\pi}, 
\label{yydiv}\\
\Pi_{yy}^{{\rm finite}}(p^2) &=& \frac{2^{[D/2]}g_D^2}{2(4\pi)^{D/2}} 2 p^2 \sum_{n=1}^\infty 
\int_0^\infty dt t^{(1-D)/2} R \sqrt{\pi} \left( 1 + \frac{(\pi Rn)^2}{t} \right) e^{-\frac{(\pi Rn)^2}{t}} \cos(2\pi na) \nonumber \\
&=&  \frac{2^{[D/2]}g_D^2}{2(4\pi)^{D/2}} 2R \sqrt{\pi} \Gamma \left(\frac{D-3}{2} \right) 
p^2 \left(\frac{D-1}{2} \right) \sum_{n=1}^\infty 
\frac{\cos(2\pi na)}{(\pi Rn)^{D-3}}. 
\label{yyfinite}
\eea
We can see that the divergence coefficients of $p^2 g_{\mu\nu} - p_\mu p_\nu$ components in $\Pi_{\mu\nu}$ 
and $p^2$ components in $\Pi_{yy}$ agree as it should be from the $D+1$ dimensional Lorentz invariance. 
That means that the ratio of the wave function renormalization factors $Z_\mu/Z_y$ are finite, 
namely the explicit finite part expression is found, 
\bea
\sqrt{\frac{Z_\mu}{Z_y}} 
&=& 1 + \frac{2^{[D/2]} g_D^2}{2(4\pi)^{D/2}} R \sqrt{\pi} \Gamma \left(\frac{D-1}{2} \right) 
\sum_{n=1}^\infty \frac{\cos(2\pi na)}{(\pi Rn)^{D-3}} \nonumber \\
&\to& 
1 + \frac{g_4^2}{16\pi^2} \sum_{n=1}^\infty \frac{\cos(2\pi na)}{n}~(D \to 4) \nonumber \\
&=& 
1 - \frac{g_4^2}{16\pi^2} \log (2\sin(\pi a))
\label{universality}
\eea 
where $\Gamma(3/2) = \sqrt{\pi}/2$. 
In the second line, we have taken the limit corresponding to the five dimensional case $D \to 4$.    
The mode sum can be carried out exactly in the last line. 
In this way, we have shown that the gauge-Yukawa universality violation at one-loop is finite 
and calculable regardless of the non-renormalizability of the model. 

\begin{figure}[ht]
 \begin{center}
  \includegraphics[width=8cm]{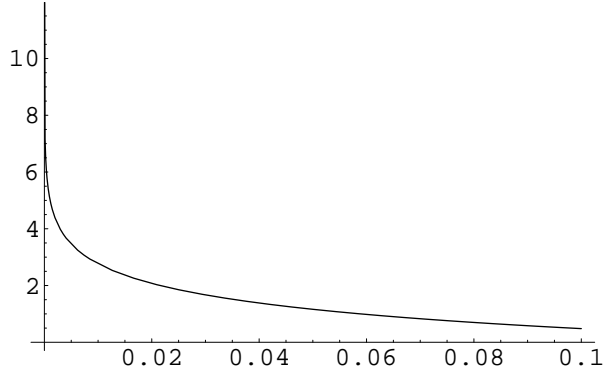}
 \end{center}
 \vspace*{-0.5cm}
\caption{
Higgs VEV dependence of the mode sum.  
The horizontal axis is Higgs VEV $a$ and the vertical axis denotes the mode sum 
$\sum_{n=1}^\infty \cos(2\pi na)/n = - \log(2\sin(\pi a))$. 
} 
\label{modesum}
\end{figure}

Next, we apply this calculable violation of gauge-Yukawa universality to 
generate a viable top quark mass. 
First we note that the ratio between top quark mass and W-boson mass can be obtained 
by multiplying the Higgs VEV to both the numerator and the denominator of the ratio (\ref{ratio}). 
Second, let us also note that the violation of the universality in (\ref{universality}) rapidly 
increases for small Higgs VEV $a$, as is shown in Fig. \ref{modesum}. 
Though we leave $a$ as a free parameter in this analysis, 
since it is highly dependent on the detail of the matter content, 
small $a$ is needed anyway to realize the electroweak symmetry breaking $SU(2)_L \times U(1)_Y \to U(1)_{{\rm em}}$ 
and sufficiently large Higgs mass.  
Because of the one-loop factor $g_4^2/(16\pi^2)$ in (\ref{universality}), 
we further need additional enhancement factor. 
In this paper, we consider the case that such an enhancement factor 
comes from group theoretical factor, {\em i.e.} 
a second rank Dynkin index $C_2(R)$ of representation $R$ 
defined as $Tr(T^a(R) T^b(R)) = C_2(R) \delta^{ab}$ 
when the fermions embedding a top quark 
belonging to large dimensional representation. 
In this case, the result (\ref{universality}) is modified only by multiplying a Dynkin index, 
\bea
\frac{m_t}{m_W} = \sqrt{\frac{Z_\mu}{Z_y}} 
&=& 
1 + \frac{g_4^2}{8\pi^2} C_2(R) \sum_{n=1}^\infty \frac{\cos(2\pi na)}{n}~(D \to 4) \nonumber \\
&=& 
1 - \frac{g_4^2}{8\pi^2} C_2(R) \log (2\sin(\pi a))
\label{top/W}
\eea
where we restricted to the case of five dimensional space-time. 

For instance, if we consider a fermion belonging to the representation with rank 4 discussed in \cite{CCP} 
to reproduce top quark mass, their Dynkin indices are given as
\bea
\begin{array}{c|cccc}
R & {\bf 15} & {\bf 24} & {\bf 27} \\
\hline
C_2(R) & 17.5 & 25 & 27 \\
\end{array}
\eea
where the normalization is taken to be $C_2(\fund) = 1/2$ for the fundamental representation. 

\begin{figure}[ht]
 \begin{center}
 \includegraphics[width=5cm]{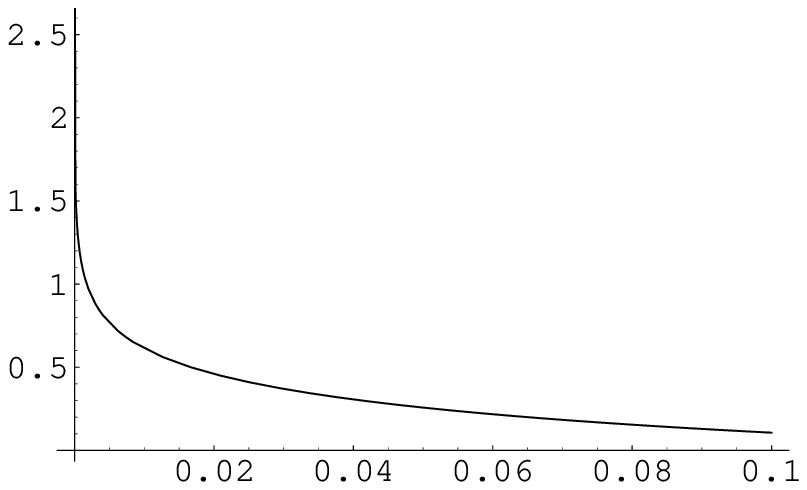}
 \hspace*{2mm}
  \includegraphics[width=5cm]{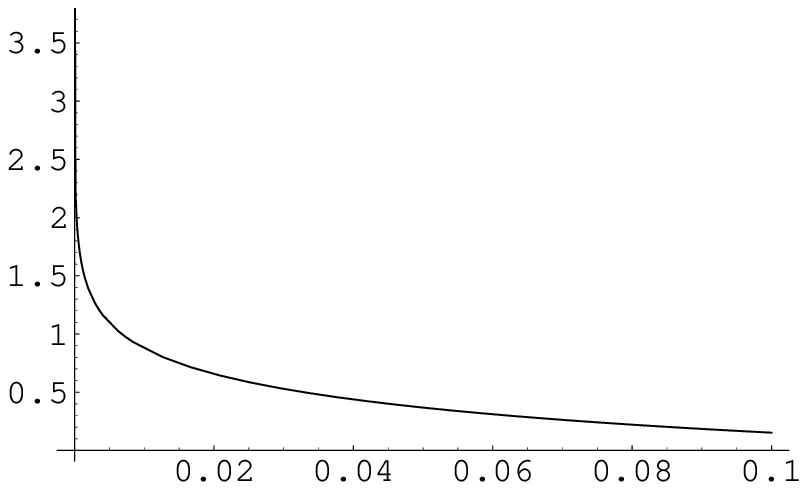}
\hspace*{2mm}
 \includegraphics[width=5cm]{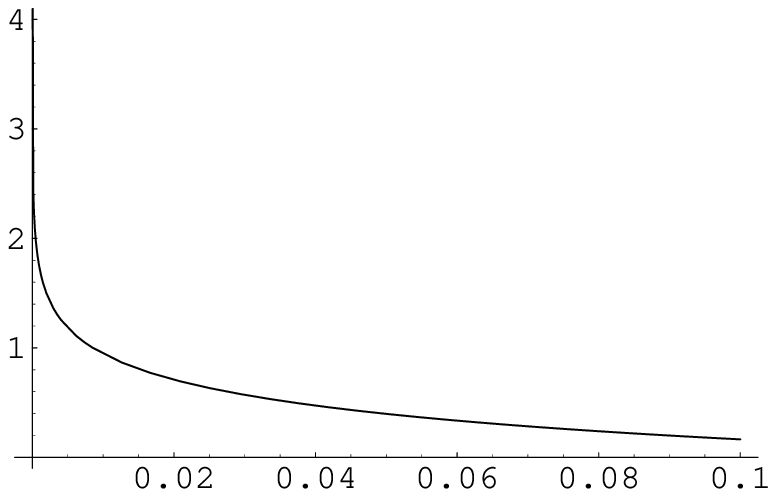}
 \end{center}
 \vspace*{-0.5cm}
\caption{
The dependence of one-loop correction $\sqrt{Z_\mu/Z_y}-1$ on $a=m_W R$ for the fermion in the representations 
${\bf 15}$ (left), ${\bf 24}$ (center) and ${\bf 27}$ (right) of $SU(3)$. 
} 
\label{KKcorrection}
\end{figure}

Now, the corresponding one-loop corrections are displayed for each representation in Fig. \ref{KKcorrection}. 
To obtain ${\cal O}(1)$ correction by compensating a factor $g_4^2/(16\pi^2)$, 
we found the upper bound on $a$ as 
$a < 0.002$ for ${\bf 15}$, $a < 0.005$ for ${\bf 24}$ and $a < 0.01$ for ${\bf 27}$. 
In other words, these constraints can be translated into those for the compactification scale 
through $m_W=a/R$, 
$R^{-1} > 40~{\rm TeV}$ for ${\bf 15}$, $R^{-1} > 16~{\rm TeV}$ for ${\bf 24}$ and $R^{-1} > 8~{\rm TeV}$ for ${\bf 27}$.

The difference of the approaches between \cite{CCP} and ours is that 
the standard model fermions are not needed to be localized at the branes in our case. 
All of the standard model fermions may be embedded in the bulk fields 
and their Yukawa coupling can be uniquely generated by the bulk gauge coupling. 
Furthermore, we do not need extra massive bulk fermions. 
This feature makes the model building (in particular the flavor sector) 
in the gauge-Higgs unification greatly simplified. 

\begin{figure}[h]
 \begin{center}
  \includegraphics[width=6cm]{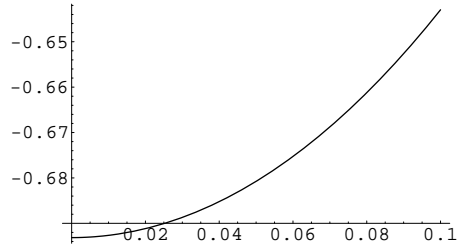}
 \end{center}
 \vspace*{-0.5cm}
\caption{
The $a$ dependence of the mode sum for the fermion with twisted boundary conditions.  
The horizontal axis is $a$ and the vertical axis denotes the mode sum $- \log(2\cos(\pi a))$. 
} 
\label{modesumtw}
\end{figure}

Next, let us consider whether the fermion with twisted boundary condition along the extra dimension 
can improve the above result. 
The contribution of the fermion with twisted boundary conditions $\Psi(y + 2\pi R) = -\Psi(y)$ can 
be straightforwardly calculated by the replacement $a\to a + \frac{1}{2}$,
\bea
\sqrt{\frac{Z_\mu}{Z_y}} &=& 1 + \frac{2^{[D/2]} g_4^2}{(4\pi)^{D/2}} C_2(R) R \sqrt{\pi} \Gamma \left(\frac{D-1}{2} \right) 
\sum_{n=1}^\infty (-1)^n \frac{\cos(2\pi na)}{(\pi Rn)^{D-3}} \nonumber \\
&\to& 
1 + \frac{g_4^2}{8\pi^2} C_2(R) \sum_{n=1}^\infty (-1)^n \frac{\cos(2\pi na)}{n}~(D \to 4) \nonumber \\
&=& 
1 - \frac{g_4^2}{8\pi^2} C_2(R) \log (2\cos(\pi a)). 
\label{twisted}
\eea 
The mode sum is shown in Fig. \ref{modesumtw}. 
We immediately see that the mode sum of twisted fermion is negative in the range $0 < a < 1$ 
and its contribution does not help to enhance the Yukawa coupling.  

In summary, we have discussed the violation of ``gauge-Yukawa universality" in the gauge-Higgs unification. 
Although the Yukawa coupling is given by the gauge coupling in the gauge-Higgs unification 
at the classical level, 
such universality is violated by quantum corrections to each coupling. 
We have shown that the violation of gauge-Yukawa universality is finite and calculable 
using in the $SU(3)$ gauge-Higgs unification model 
with arbitrary space-time dimension compactified on an orbifold $S^1/Z_2$. 
The point is that the gauge-Yukawa universality violation is parameterized 
by the ratio of Yukawa coupling and the gauge coupling and 
the ratio is further expressed by the ratio of the wave function renormalization factor for the gauge field and 
the extra component of the gauge field. 
The ratio is clearly understood to be finite from the higher dimensional Lorentz invariance. 

As an interesting application, we have proposed a mechanism 
to generate a viable top quark mass in flat space gauge-Higgs unification, 
alternative to the mechanism in \cite{CCP}. 
By multiplying the Higgs VEV, 
the violation of gauge-Yukawa universality can lead the quantum correction to the top quark mass. 
We have shown that the order one correction to the top Yukawa coupling is possible 
if the fermion belongs to the large dimensional representation and the Higgs VEV is very small 
compared with the compactification radius, {\em i.e.} $a=M_W R \ll 1$. 
As a result, we have obtained the constraints for the compactification scale 
in the case where fermions belong to the rank 4 representation 
(${\bf 15}$, ${\bf 24}$,  and ${\bf 27}$ representations of $SU(3)$) 
discussed in the literature \cite{CCP}. 
One of the advantages of our approach is 
that the standard model fermion is not needed to be localized on the branes 
and no extra massive bulk fermions are required. 
This makes the model building of the gauge-Higgs unification (in particular the flavor physics) greatly simplified. 

We hope that this approach shed some new insights on the flavor physics of the gauge-Higgs unification. 

\subsection*{Acknowledgments}
The work of the authors was supported 
in part by the Grant-in-Aid for Scientific Research 
of the Ministry of Education, Science and Culture, No.18204024 and No. 20025005.  

\end{document}